# Chemical Reactions under Nanoconfinement:
# Unravelling Equilibrium Constant Equations

Leonid Rubinovich and Micha Polak

*Department of Chemistry, Ben-Gurion University of the Negev, Beer-Sheva, Israel*

Equilibrium Constant Differential Equations (ECDE) are derived for several nanoconfined elemental bimolecular reactions in the frameworks of statistical mechanics and the ideal gas model. The ECDEs complement the well-known equilibrium-constant ordinary equations that are used for macroscopic systems. Solving the ECDE numerically or analytically furnishes the average reaction extent, as well as its variance and skewness. This original theoretical-computational methodology fills the gap in studies of nanochemical equilibrium providing a consistent and convenient alternative to derivations based on direct employment of the canonical partition-functions. Whereas the latter become more complex and time-consuming with increased number of molecules, the ECDE-based computations are equally efficient for small as well as large numbers of nanoconfined reacting molecules. The ECDE methodology introduced here is confirmed by a complete agreement with partition-function computations. In addition, the new approach is applied to nanoconfined adsorption.

## 1. Equilibrium Constant Equations

Chemical-equilibrium involving a small number of molecules inside a confined nanospace can exhibit considerable deviations from the macroscopic thermodynamic limit (TL) due to reduced mixing entropy, as was predicted in several of our works using statistical-mechanics canonical partition-functions and the ideal-gas model[1,2]. For example, exergonic addition and dimerization should exhibit a considerable shift of the reaction extent towards product formation[3]. This "nanoconfinement entropic effect on chemical-equilibrium" (NCECE) was verified by post-analysis[4] of reported measurements of DNA hybridization inside confined nano-fabricated chambers[5] as well as by stochastic kinetics modeling[6,7].

The present work is a substantial extension of the NCECE studies, focusing now on variations of the ordinary equilibrium constant equations (ECE) due to nanoconfinement. Equilibrium elementary chemical reaction is considered,

$$\sum r_i R_i \leftrightarrow \sum p_j P_j, \tag{1.1}$$

where $R_i$ (or $P_j$) and $r_i$ (or $p_j$) denote reactants (or products) and the corresponding stoichiometric coefficients, respectively. In this study, the frequently made assumption is made that all activity coefficients are close to 1 (ideal system), so the value of the thermodynamic equilibrium constant, $K$, equals to the constant in terms of molar concentrations[8]. Namely, in the TL,

$$K = (N_{Av} V)^{\sum r_i - \sum p_j} \frac{\left(\prod N_j^{p_j}\right)_{TL}}{\left(\prod N_i^{r_i}\right)_{TL}}. \tag{1.2}$$

where $N_i$ denotes the equilibrium number of molecules $i$ ($N_{Av}$ is Avogadro's number, and the volume, $V$, is given in liters).

Since the numbers of product molecules are related to the reaction extent, its value can be found by solving equations simply derived from Eq.1.2. However, this ECE cannot be used for nanoconfined reactions, since its RHS differs from $K$ and equals to a distinct "nanoequilibrium constant"[1] (sometimes called "apparent equilibrium constant"). The main goal of the present study is to find ECEs, which substitute the conventional ones in the case of nanoconfinement. Solutions of the latter should provide the average nanoreaction extent, $\xi$, as a function of the thermodynamic equilibrium constant, $K$.

It can be noted that each molecule involved in the reaction equation 1.1 contributes to $K$ a multiplier that is equal to the number of such molecules, which drops (or rises) by $r_i$ (or $p_j$) for reactants (or products) in a single reaction step. Since in the TL $N_i$ is huge, its alteration is negligible and equation 1.2 includes the factor $N_i^{r_i}$ for reactant $R_i$. However, in the case of a limited nanoconfined molecule reservoir this factor should be substituted by 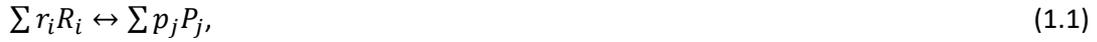 $N_i(N_i - 1)(N_i - 2) \ldots (N_i - r_i + 1) = \frac{N_i!}{(N_i - r_i)!}$. Likewise, for the products $N_j^{p_j}$ should be substituted by $N_j(N_j + 1)(N_j + 2) \ldots (N_j + p_j - 1) = \frac{(N_j + p_j - 1)!}{(N_j - 1)!}$. Correspondingly, the equilibrium constant can be generalized as,

$$K = (N_{Av} V)^{\sum r_i - \sum p_j} \frac{\left\langle \prod \frac{(N_j + p_j - 1)!}{(N_j - 1)!} \right\rangle}{\left\langle \prod \frac{N_i!}{(N_i - r_i)!} \right\rangle}. \tag{1.3}$$

where the canonical ensemble average, $\langle ... \rangle$, is used for fluctuating nanosystems. In the TL $\frac{N_i!}{(N_i-r_i)!} \to N_i^{r_i}$ and $\frac{(N_j+p_j-1)!}{(N_j-1)!} \to N_j^{p_j}$ and therefore the original Eq.1.3 is consistent with the conventional definition.

The validity of the generalized Eq.1.3 is confirmed by calculations below for several nanoconfined chemical reactions. It is the basis for novel Equilibrium Constant Differential Equations (ECDE) equations for $\xi$ as a function of $K$, distinctly from the conventional ECEs in the TL case.

## 2. Derivation of the ECDEs

Starting with the exchange reaction $A + B = C + D$, the numbers of reactant molecules are related to the "reaction extent", $x$, as $N_A = N(1 - x)$ and $N_B = N(r - x)$, and the numbers of product molecules are equal to $N_C = N_D = Nx$ ($r \geq 1$, the equality holds in the stoichiometric case). The initial (maximal) numbers of reactants are $N_A^{(0)} = N$ and $N_B^{(0)} = rN$. According to Eq.1.3 the equilibrium constant is given by,

$$K = \frac{\langle N_C N_D \rangle}{\langle N_A N_B \rangle} = \frac{\langle x^2 \rangle}{r - r\langle x \rangle - \langle x \rangle + \langle x^2 \rangle}. \tag{2.1}$$

For $2A = B + C$ with $N_A = 2N(r - x)$ and $N_B = N_C = Nx$, where $r = 1$ and $r = 1 + \frac{1}{2N}$, if $N_A^{(0)}$ is even and odd, respectively. The equilibrium constant reads,

$$K = \frac{\langle N_B N_C \rangle}{\langle N_A(N_A-1) \rangle} = \frac{\langle x^2 \rangle}{4\langle (r-x)^2 - \frac{1}{2N}(r-x) \rangle}. \tag{2.2}$$

In the case of the addition reaction $A + B = C$ with $N_A = N(1 - x)$, $N_B = N(r - x)$ and $N_C = Nx$ the equilibrium constant is given by,

$$K = N_{Av} V \frac{\langle N_C \rangle}{\langle N_A N_B \rangle} = N_{Av} V \frac{\langle x \rangle}{N\langle (1-x)(r-x) \rangle} = \frac{\langle x \rangle}{[A]^{(0)}(r^2 - 2r\langle x \rangle + \langle x^2 \rangle)}, \tag{2.3}$$

where $[A]^{(0)}$ denotes the initial (maximal) molarity of the reactant.

For the dimerization reaction $2A = B$ with $N_A = 2N(r - x)$ and $N_B = Nx$, where $r = 1$ and $r = 1 + \frac{1}{2N}$ if $N_A^{(0)}$ is even and odd, respectively. The equilibrium constant is,

$$K = N_{Av} V \frac{\langle N_B \rangle}{\langle N_A(N_A-1) \rangle} = N_{Av} V \frac{\langle x \rangle}{\langle 4N(r-x)^2 - 2(r-x) \rangle} = \frac{r\langle x \rangle}{2[A]^{(0)}[r^2 - 2r\langle x \rangle + \langle x^2 \rangle - \frac{1}{2N}(r - \langle x \rangle)]}, \tag{2.4}$$

where $[A]^{(0)}$ denotes the initial (maximal) molarity of the reactant.

By substitution of $\xi \equiv \langle x \rangle$ and $\langle x^2 \rangle = \xi^2 + \frac{1}{N} \frac{\partial \xi}{\partial \ln K}$ to Eqs.2.1-4 (to be published elsewhere), original ECDEs are obtained, for $A + B = C + D$, $\frac{1}{N}(K - 1) \frac{\partial \xi}{\partial \ln K} = \xi^2 - K(1 - \xi)(r - \xi),$ (2.5) for $2A = B + C$, $\frac{1}{N}(4K - 1) \frac{\partial \xi}{\partial \ln K} - \frac{2K}{N}(r - \xi) = \xi^2 - 4K(r - \xi)^2,$ (2.6) for $A + B = C$,  $\quad \frac{K}{N}\frac{\partial \xi}{\partial lnK} = \frac{\xi}{[A]^{(0)}} - K(1-\xi)(r-\xi)$,  (2.7)

and

for $2A = B$,  $\quad \frac{K}{N}\left(\frac{\partial \xi}{\partial lnK} - \frac{1}{2}(r-\xi)\right) = \frac{r\xi}{2[A]^{(0)}} - K(r-\xi)^2$.  (2.8)

In addition to numerical solutions provided e.g., by MATLAB for the derived ECDEs, analytical solutions exist in terms of Jacobi and Laguerre polynomials[9] (Table 1). Since the LHS in Eqs.2.5-8 is on the order of $\frac{1}{N}$, it approaches zero when $N$ is large, thereby the ordinary ECEs are obtained in the TL. Furthermore, they become reasonably accurate even for $N \geq 10$, consistently with the decreasing NCECE effects predicted earlier for this condition[1].

It can be shown that differentiation of $\xi$ with respect to $lnK$ provides characteristics of the reaction-extent such as the variance (fluctuations),

$\sigma^2 \equiv \langle x^2 \rangle - \xi^2 = \frac{1}{N}\frac{\partial \xi}{\partial lnK}$,  (2.9)

and its skewness (the distribution asymmetry),

$Sk \equiv \frac{\langle (x-\xi)^3 \rangle}{\sigma^3} = N^{-1/2}\left(\frac{\partial \xi}{\partial lnK}\right)^{-3/2}\frac{\partial^2 \xi}{\partial (lnK)^2}$,  (2.10)

As an example, the exchange reaction $A + B = C + D$ is chosen for further analysis. Using the analytical solution for the stoichiometric case given in Table 1 and the asymptotic approximation $\lim_{x \to 1} P_n^{(\nu,\mu)}(x) = \frac{\Gamma(n+\nu+1)}{\Gamma(n+1)\Gamma(\nu+1)}$ [10], one obtains $\xi \sim 1 - \frac{N}{K}$ for large $K$. Thus, the reaction is closer to completion when the number of molecules is small, consistently with the ordinary NCECE. On the other hand, $\xi \sim NK$ for small $K$, i.e. the average reaction extent diminishes for smaller numbers of molecules (inverse NCECE). Plots of the average reaction extent as a function of $lnK$ reflect distinctly the increasing NCECE effects in smaller size systems (Fig. 1a). As can be seen, $\xi$ exhibits enhancement in the exergonic case and diminution in the endergonic case. The maximal variance at $\xi \approx 0.5$ (and $lnK \approx 0$) corresponds to high-$T$ widening of a quite uniform microstate-probability distribution, which consistently narrows with temperature decrease and for larger system sizes (Fig. 1b). Furthermore, NCECE effects are related to the distribution asymmetry, which is right-skewed for $\xi < 0.5$ and left-skewed for $\xi > 0.5$ (Fig. 1b). Similarly to the variance, the skewness values decrease for larger system sizes. The validity of the derivations and computations based on the present original methodology is confirmed by complete agreement with predictions obtained by direct employing of canonical partition-functions.

The final derivation concerns non-dissociative adsorption under nanoconfinement having an apparent formal similarity with the addition reaction considered above, namely $G + A^* = A$, where $G$ and $A^*$ denote a gas molecule and a surface site, respectively. Focusing on the stoichiometric case ($r = 1$), just a slight modification is required in terms of the equilibrium average coverage, $\theta$ (substituting $\xi$), and the initial (maximal, pre-adsorption) ideal-gas pressure, $p^{(0)} = 10^3 N_{Av} kT[G]^{(0)}$. In the TL,

$K = N_{Av}V \frac{(N_A)_{TL}}{(N_G N_{A^*})_{TL}} = N_{Av}kTK_L$,  (2.11)

where $K_L = \frac{\theta_{TL}}{p(1-\theta_{TL})}$ is the "Langmuir adsorption constant". So, the original differential equation for the average nanoconfined coverage as a function of the Langmuir constant is obtained by substitution of $[G]^{(0)}K$ by $p^{(0)}K_L$ in Eq.2.7,

$$\frac{K_L}{N}\frac{\partial \theta}{\partial \ln K_L} = \frac{\theta}{p^{(0)}} - K_L(1-\theta)^2. \tag{2.13}$$

Consistently with the addition reaction solution (Table 1), the analytical solution of equation (2.13) is given by,

$$\theta = 2 + \frac{1}{p^{(0)}K_L} + \frac{1}{N}\left(1 - (N+1)\frac{L_{N+1}\left(-\frac{N}{p^{(0)}K_L}\right)}{L_N\left(-\frac{N}{p^{(0)}K_L}\right)}\right). \tag{2.14}$$

In the TL ($N \to \infty$) the LHS of Eq.2.13 equals zero and as $p = p^{(0)}(1-\theta)$ the Langmuir isotherm is obtained,

$$\theta = \frac{pK_L}{1+pK_L}. \tag{2.15}$$

### *3. Conclusions*

Equilibrium-constant differential equations (ECDEs) are derived for several binary chemical reactions and for adsorption under nanoconfinement. When the system size increases approaching the TL the ECDEs gradually evolve to the ordinary ECEs. The ECDEs fill the gap in studies of nanochemical equilibrium providing a consistent and convenient alternative to the more common method directly employing canonical partition-functions. Whereas the latter becomes more complex and time consuming for larger systems, the original theoretical-computational methodology employing the ECDEs is equally efficient for nanosystems containing small and large numbers of molecules. The validity of the ECDE-methodology introduced here is confirmed by the complete agreement of computation results with those based on the partition-function computations. Furthermore, the results of numerical solutions of the ECDEs provided by MATLAB package coincide with those obtained by previously reported analytical solutions obtained by cumbersome stochastic-kinetic modeling for long times. Applications of the ECDE method to termolecular reactions, such as $H_2 + 2I = 2HI$, are underway. According to preliminary results, the termolecular NCECE effect should be more significant compared to binary reactions.

Table 1. Equilibrium constant differential equations (ECDEs) and their analytical solutions for several elementary binary reactions

| Reaction | ECDE* | Analytical solution** |
|---|---|---|
| $A + B = C + D$<br>$N_A^{(0)} = N$<br>$N_B^{(0)} = rN$ | $\frac{1}{N}(K-1)\frac{\partial \xi}{\partial \ln K} = \xi^2 - K(1-\xi)(r-\xi)$ | $\xi = 1 - \dfrac{P_{N-1}^{(Nr-N+1,-Nr-N)}\left(1-\frac{2}{K}\right)}{K P_N^{(Nr-N,-Nr-N-1)}\left(1-\frac{2}{K}\right)}$ |
| $2A = B + C$<br>$N_A^{(0)} = 2rN$<br>Even $N_A: r = 1$<br>Odd $N_A: r = 1 + \frac{1}{2N}$ | $\frac{1}{N}(4K-1)\frac{\partial \xi}{\partial \ln K} - \frac{2K}{N}(r-\xi) = \xi^2 - 4K(r-\xi)^2$ | Even $N_A: \xi = 1 - \dfrac{P_{N-1}^{(\frac{1}{2},-2N+\frac{1}{2})}\left(1-\frac{1}{K}\right)}{2K P_N^{(-\frac{1}{2},-2N-\frac{1}{2})}\left(1-\frac{1}{K}\right)}$<br><br>Odd $N_A: \xi = 1 - \dfrac{3 P_{N-1}^{(\frac{3}{2},-2N-\frac{1}{2})}\left(1-\frac{1}{K}\right)}{2K P_N^{(\frac{1}{2},-2N-\frac{3}{2})} P_N^{(\frac{1}{2},-2N-\frac{3}{2})}\left(1-\frac{1}{K}\right)}$ |
| $A + B = C$<br>$N_A^{(0)} = N$<br>$N_B^{(0)} = rN$ | $\frac{K}{N}\frac{\partial \xi}{\partial \ln K} = \frac{\xi}{[A]^{(0)}} - K(1-\xi)(r-\xi)$ | $\xi = 1 + r + \frac{1}{[A]^{(0)} K} + \frac{1}{N}\left(1 - (N+1)\dfrac{L_{N+1}^{(Nr-N)}\left(-\frac{N}{[A]^{(0)}K}\right)}{L_N^{(Nr-N)}\left(-\frac{N}{[A]^{(0)}K}\right)}\right)$ |
| $2A = B$<br>Even $N_A: r = 1$<br>Odd $N_A: r = 1 + \frac{1}{2N}$ | $\frac{K}{N}\left(\frac{\partial \xi}{\partial \ln K} - \frac{1}{2}(r-\xi)\right) = \frac{r\xi}{2[A]^{(0)}} - K(r-\xi)^2$ | Even $N_A: \xi = 1 - \dfrac{N}{2[A]^{(0)}K} \dfrac{L_{N-1}^{(\frac{1}{2})}\left(-\frac{N}{2[A]^{(0)}K}\right)}{L_N^{(-\frac{1}{2})}\left(-\frac{N}{2[A]^{(0)}K}\right)}$<br><br>Odd $N_A: \xi = 1 - \dfrac{rN}{2[A]^{(0)}K} \dfrac{L_{N-1}^{(\frac{3}{2})}\left(-\frac{rN}{2[A]^{(0)}K}\right)}{L_N^{(\frac{1}{2})}\left(-\frac{rN}{2[A]^{(0)}K}\right)}$ |

\* The well-known ordinary equilibrium constant equations in the TL (ECEs) are obtained when $N \to \infty$ so LHS$\to 0$.

\*\* $P_n^{(\nu,\mu)}(x)$ and $L_n^{(m)}(x)$ denote the Jacobi and the associated Laguerre polynomials, respectively[9]. The analytical solutions are consistent with those obtained by stochastic-kinetic modeling for long times that involve hypergeometric functions[11].

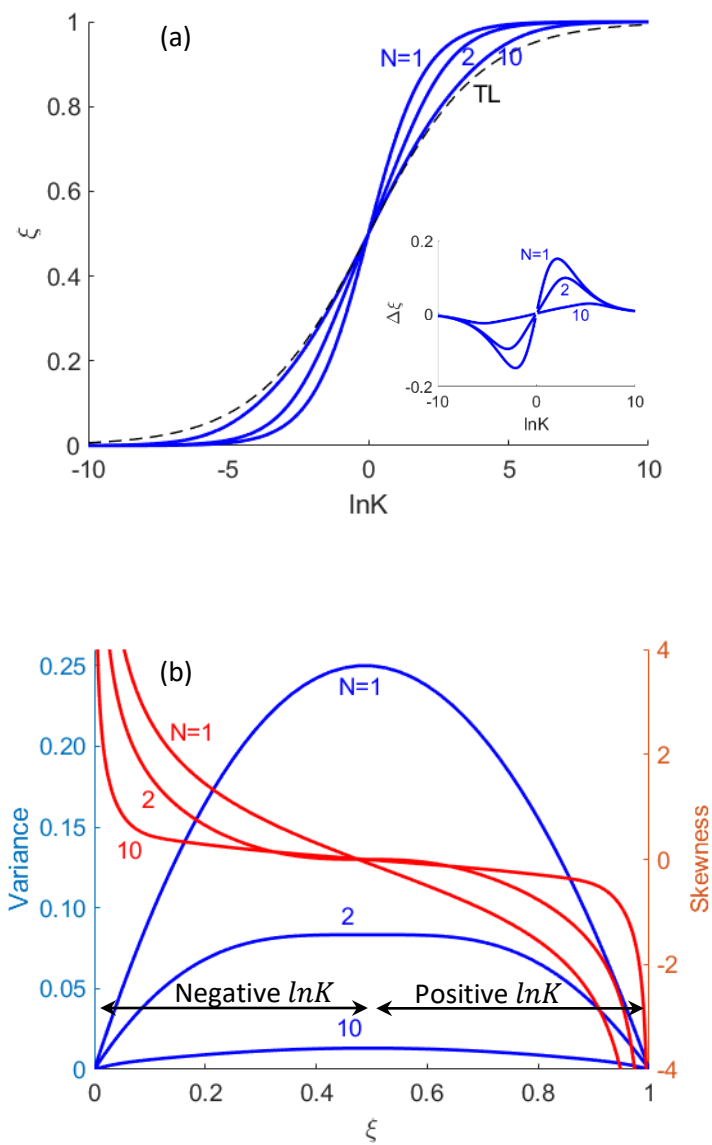

Fig. 1. The NCECE effects for exergonic and endergonic $A + B = C + D$ exchange reactions. (a) Equilibrium extents versus $lnK$, inset: $\Delta\xi \equiv \xi - \xi^{TL}$ (b) Variance versus $\xi$ for the stoichiometric reaction (blue lines) computed for the marked initial number of reactant molecule pairs. The corresponding skewness of the reaction extent is given by red lines. The values of $lnK$ can be considered as linearly related to (i) the inverse temperatures of a given reaction, or to (ii) energies of different reactions at a given temperature.